\documentclass{emulateapj}

\usepackage{natbib}
\usepackage{psfig}
\usepackage{graphicx}

\shorttitle{Stellar X-ray point source sensitivity}
\shortauthors{Wright et al.}

\begin{document}

\title{Simulating the sensitivity to stellar point sources of {\it Chandra} X-ray observations}

\author{Nicholas J. Wright$^{1,2}$, Jeremy J. Drake$^2$, Mario G. Guarcello$^{2,3}$, Vinay L. Kashyap$^2$, Andreas Zezas$^{2,4}$}
\affil{$^1$Centre for Astrophysics Research, University of Hertfordshire, Hatfield AL10 9AB, UK}
\email{nick.nwright@gmail.com}
\affil{$^2$Center for Astrophysics, 60 Garden Street, Cambridge, MA~02138, USA}
\affil{$^3$INAF - Osservatorio Astronomico di Palermo, Piazza del Parlamento 1, I-90134 Palermo, Italy}
\affil{$^4$Physics Department, University of Crete, GR-710 03 Heraklion, Crete, Greece}

\begin{abstract}

The {\it Chandra} Cygnus OB2 Legacy Survey is a wide and deep X-ray survey of the nearby and massive Cygnus~OB2 association. The survey has detected $\sim$8,000 X-ray sources, the majority of which are pre-main sequence X-ray emitting young stars in the association itself. To facilitate quantitative scientific studies of these sources as well as the underlying OB association it is important to understand the sensitivity of the observations and the level of completeness the observations have obtained. Here we describe the use of a hierarchical Monte Carlo simulation to achieve this goal by combining the empirical properties of the observations, analytic estimates of the source verification process, and an extensive set of source detection simulations. We find that our survey reaches a 90\% completeness level for a pre-main-sequence population at the distance of Cyg~OB2 at an X-ray luminosity of $4 \times 10^{30}$~ergs~s$^{-1}$ and a stellar mass of 1.3~M$_\odot$ for a randomly distributed population. For a spatially clustered population such as Cyg~OB2 the 90\% completeness level is reached at 1.1~M$_\odot$ instead, as the sources are more concentrated in areas of our survey with a high exposure. These simulations can easily be adapted for use with other X-ray observations and surveys, and we provide X-ray detection efficiency curves for a very wide array of source and background properties to allow these simulations to be easily exploited by other users.

\end{abstract}


\section{Introduction}

To maximise the scientific potential of any catalog of sources it is important to characterise the sensitivity of the observations to sources with given properties. This allows the user of a catalog to understand the strengths and limits of the observations and the reduced data set. This is particularly important for X-ray observations, such as those using the {\it Chandra X-ray Observatory} \citep{weis02}, where the sensitivity is not uniform across the field of view due to a combination of vignetting and the variable size and shape of the point spread function (PSF). The latter is particularly important as it largely determines the background count rate that limits source detection and significance.

For the {\it Chandra} Cygnus OB2 Legacy Survey \citep{drak15,wrig14c,guar15}, which is devoted to uncovering the pre-main-sequence population of Cyg~OB2, the largest group of young stars within 2~kpc of the Sun \citep{mass91,hans03}, this issue is further complicated by the observational tiling strategy adopted \citep[see Figure~1 of][]{wrig14c}. Compared to other X-ray surveys, where either different pointings do not overlap considerably \citep[e.g.,][]{wrig09a,guar12b}, or the pointings are co-axial but vary in roll angle \citep[e.g.,][]{gunt12}, the different observations in our survey are both heavily overlapping and not co-axial. This leads to different PSFs and vignetting factors for each observation of each source in the survey, making the problem of assessing the sensitivity of the survey much more complex than is often the case.

A common method of quantifying the sensitivity of a given set of observations is to simulate the detection procedure by inserting false sources into the observations and then subjecting them to the same source detection and verification procedure used on the actual sources. While this method has many advantages it would be very difficult and time consuming to implement on a complex data set such as ours, and we therefore concluded that an alternative approach would be necessary. To evaluate the sensitivity of our survey we have developed a hierarchical Monte Carlo simulation that combines simulations of the source detection and verification process, the empirical properties of our observations, and a model of the stellar X-ray sources the survey is targeting.

It is important to be careful with language when discussing the sensitivity or completeness of a survey or source catalog. In this paper we use the word ``completeness'' to mean, for a given set of observations, the probability of detecting a source as a function of some property of that source. That property may be either an observational property, such as the measured flux from that source that depends on its distance from us, or, by means of some assumptions about the sources, a property that is inherent to the source itself, such as its luminosity or its mass. We also use the term completeness to mean the probability that a member of a population of sources (such as the young stars in Cyg~OB2) will be included within a given source catalog. A level of completeness can be determined for a source catalog and a population of sources by making a number of assumptions about that population, such as their spatial distribution and their distribution as a function of one of the parameters to which the catalog's completeness is known.

This paper is arranged as follows. In Section~2 we outline the methodology used for our source detection and verification simulations, provide a quantification of the source detection process over a range of different parameters, and describe the physical model that predicts the X-ray photons that {\it Chandra} should see for various stellar sources. Then in Section~3 we present and discuss the results of our simulations, giving the completeness of our X-ray survey as a function of various quantities such as source count rate, the stellar X-ray luminosity, and the stellar mass.

\section{Methodology}

The objective of this work is to quantify the completeness of our survey catalog as a function of various observational (X-ray count rate) and stellar (X-ray luminosity and stellar mass) parameters. To achieve this the approach adopted here is to perform a Monte Carlo simulation of the source detection and verification process across our entire survey area using the intrinsic and empirical properties of our observations. In this section we outline the components of our Monte Carlo simulation and the data used to perform each part of it.

The process begins by simulating the intrinsic and observed properties of the population we wish to quantify the detection of, which in this case is a pre-main-sequence (PMS) stellar population of X-ray emitters, affected by interstellar hydrogen absorption (Section~\ref{s-sources}). These sources are then randomly distributed over the entire survey area and based on the position of the source we know from the actual survey grid the number of times that source would have been observed in our observations and the exposure time of each observation. We can also extract from the actual observations the empirical X-ray background level at that position in each observation (Section~\ref{s-observations}). Using this information we then assess the probability that the source would be detected in our observations using the source detection strategy employed in producing our X-ray catalog (Section~\ref{s-detection}), which we base on an extensive suite of source detection simulations we have carried out. Finally we use the same information to simulate the source verification process that candidate sources in our catalog were put through (Section~\ref{s-verification}). Once complete the results of our Monte Carlo simulation provide the fraction of sources detected as a function of a given observational or intrinsic source property.

\subsection{The properties of stellar X-ray sources}
\label{s-sources}

The ultimate objective of this work is to quantify the completeness of our survey as a function of stellar mass for a population of PMS stellar X-ray sources. This is desirable because studies of the age \citep{wrig10a} or structure \citep{wrig14b} of Cyg~OB2, or studies of the evolution of the disk-bearing stars in the association \citep{wrig12a,guar15b}, are usually performed as a function of stellar mass. As intermediate results we will also be able to quantify the survey completeness as a function of X-ray count rate and luminosity.

\subsubsection{The stellar mass versus X-ray luminosity relation}

To simulate the X-ray emission from stars of a given mass we use the relationship between stellar mass and X-ray luminosity quantified by \citet{tell07} from studies of young stars in the Taurus molecular cloud. We used this relationship because it is in good agreement with that from other studies \citep[e.g.,][]{prei05a}, is believed to be complete over the mass range of interest (0.5--3.0~M$_\odot$), and should not be biased by X-ray luminosity as the authors note that the majority of targets were detected well above their detection limit. While Cyg~OB2 contains many stars more massive than this \citep[e.g.,][]{wrig15a}, the most massive O-type stars are known to be detected with 100\% efficiency \citep{rauw14} while the intermediate-mass A- and B-type stars are known to be X-ray `dark' \citep[e.g.,][]{drak14a} and therefore are often not detected.

\citet{tell07} quantify relationships between $L_X$ and mass for both classical T-Tauri stars (CTTS, log~$L_X = 1.98 \, \mathrm{log} \, M + 30.24$) and weak-lined T-Tauri stars (WTTS, log~$L_X = 2.08 \, \mathrm{log} \, M + 30.69$), the X-ray luminosities of which are known to differ significantly \citep{prei05a}. We assign stars in our Monte Carlo simulation as either CTTS or WTTS on the assumption that $\sim$5\% of stars in the association are accreting \citep[][found 10 CTTS out of $\sim$250 spectroscopically-observed stars in Cyg~OB2, see also \citealt{guar13}]{vink08}, and assign them X-ray luminosities according to the relevant relationship, adding in log normal dispersions of 0.45 (CTTS) or 0.38~dex (WTTS) as noted by \citet{tell07}. 

Since the stars in Taurus are likely to be slightly younger than those in Cyg~OB2 \citep[typical ages of 1--3~Myr compared to 3--5~Myr,][]{wrig10a} we also use the relationship between age and $L_X$ found by \citet{tell07}, which has a slope of $-0.36 \, \mathrm{log} \, (\tau / Myr)$~dex. We therefore randomly assign stars an age in the range 3--5~Myr \citep{wrig10a} and correct their X-ray luminosities appropriately.

It is worth noting that the X-ray luminosities of low- and solar-mass stars are not constant but are characterised by many short-duration flare-like events, which our simple stellar X-ray model has not taken into account. \citet{tell07} excluded the largest flares from the observed light-curves when calculating the stellar mass to X-ray luminosity relationships that we have used. This will mean that for a small fraction of stars in our simulation the $L_X$ values we have calculated will be underestimated and therefore our completeness at a given mass will also be underestimated. We experimented with adding a simple flare model to take into account the variation in $L_X$ induced by short-duration flares, but found that it only introduced a small difference in the observed X-ray luminosity distribution, $<10$\%. Furthermore, because we do not know exactly how many and what size of flares were excluded by \citet{tell07}, we cannot accurately include this in our model; however the magnitude of the effect appears to be small.

\subsubsection{Conversion from X-ray luminosity to count rate}

To convert from observed X-ray luminosity to X-ray count rate in one of {\it Chandra}'s detectors we must assume three quantities: the distance to Cyg~OB2, the plasma temperature of the X-ray emitting sources (which determines the X-ray spectrum emitted), and the line-of-sight absorbing hydrogen column density (which influences the X-ray spectrum observed). These three quantities are generally sufficient for simulating the X-ray spectrum that {\it Chandra} observes, which can then be converted into a measured count rate (in a given band) using {\it Chandra}'s auxiliary response files (ARF\footnote{The ARF contains the combined telescope, filter and detector effective areas and quantum efficiencies as a function of energy and averaged over time. When an input spectrum is multipled by the ARF the result is the distribution of counts seen by a detector with perfect energy resolution.}) and redistribution matrix files (RMF\footnote{The RMF reproduces the spread in photon energy (measured by {\it Chandra} as detector pulse height) that {\it Chandra} measures for an observed photon with a given energy.}). Since we are only interested in calculating broad band count rates and not X-ray spectra the use of RMFs is not vital. ARFs are important however because they include variations in the effective area, e.g., due to vignetting, over the ACIS-I\footnote{The Advanced CCD Imaging Spectrometer \citep{garm03}.} CCD. Despite this, it would be too time consuming to simulate each individual source spectrum, apply the absorption due to neutral hydrogen, multiply by the ARF at the position of the source on the detector, and then integrate the number of events that would be detected.

To simplify this situation we instead use {\it Chandra}'s Portable Interactive Multi-Mission Simulator\footnote{PIMMS, http://cxc.harvard.edu/toolkit/pimms.jsp.} to calculate the number of counts detected by the ACIS-I detector as a function of the plasma temperature and flux of the source spectrum, and the column density of neutral hydrogen along the line of sight. Since PIMMS uses a single value of the ARF for these calculations we also introduce a spatially-varying correction factor to reproduce the variation of the ARF over all our observations. Since the spectral shape of the ARF does not change significantly, even as the magnitude of the ARF varies\footnote{The shape of the ARF over the area of the ACIS-I CCD varies by at most 10\% over the energy range 1--5~keV, while outside of this range the variation increases to $\sim$20\%, but both the ARF and our input spectra are greatly reduced and so the impact of this variation on our results is greatly reduced.}, we implement this correction factor as the ratio of the actual ARF (extracted from our observations) and the ARF used by PIMMs, both calculated at an energy of 1.49~keV, the approximate peak of the ARF. By implementing this ARF correction we can account for the variation of the ACIS-I effective area over the CCD without slowing down our simulations. We find that this correction factor influences our final completeness fractions at a given flux or stellar mass by $\sim$5--10\%.

We use a distance to Cyg~OB2 of $1.40 \pm 0.08$~kpc, a value that is well constrained from multiple studies including parallax measurements to Cyg~X \citep{rygl12}. For the plasma temperature we assume that our sources are well characterised by a single temperature thermal plasma and randomly assign our sources a plasma temperature sampled from those measured in the Orion Nebula Cluster \citep{getm05}, which show a clear peak at $\sim$0.8~keV, with a high energy tail extending to $\sim$4~keV. We chose this distribution because the sample size is large and it originates from a well understood and low-extinction population. Some studies have found that stellar X-ray spectra are better fit by a two-temperature thermal plasma, though this does not result in a large difference to the modelled spectrum, especially for low counts data, and therefore we have not not used this approach.

For the absorbing column of neutral hydrogen we use the distribution of measured visual extinction found by \citet{wrig15a}, which we convert into an absorbing hydrogen column density using the commonly used relation $N_H = 2.2 \times 10^{21} A_V$~cm$^{-2}$ \citep{ryte96}. The distribution is well approximated by a Gaussian centered at log~$N_H = 22.09$~cm$^{-2}$ and with $\sigma = 0.095$~dex. In the center of the association this distribution is in good agreement with that found by \citet{alba07} from their X-ray study of the core of Cyg~OB2, but better represents the full range of extinctions observed across the entire association.

Using these quantities and assuming a solar-metallicity thermal X-ray spectrum \citep{raym77} we use PIMMS to calculate count rates in each of the three energy bands in our survey: {\it broad} (0.5--7~keV), {\it soft} (0.2--2~keV), and {\it hard} (2--7~keV).

\subsection{Simulating the observational grid and empirical background levels}
\label{s-observations}

A key element of our Monte Carlo simulation is that it uses the intrinsic properties of our observations (the positions, roll angles and areas of each observation) and the empirical properties of the X-ray background in each observation and at each position. Based on the randomly determined position of each simulated source we use the field of view of each observation used in our survey \citep{wrig14c} to determine which ObsIDs would have observed the source, the off-axis angle of the source, and the exposure time of the observation.

\begin{figure*}
\begin{center}
\includegraphics[width=450pt]{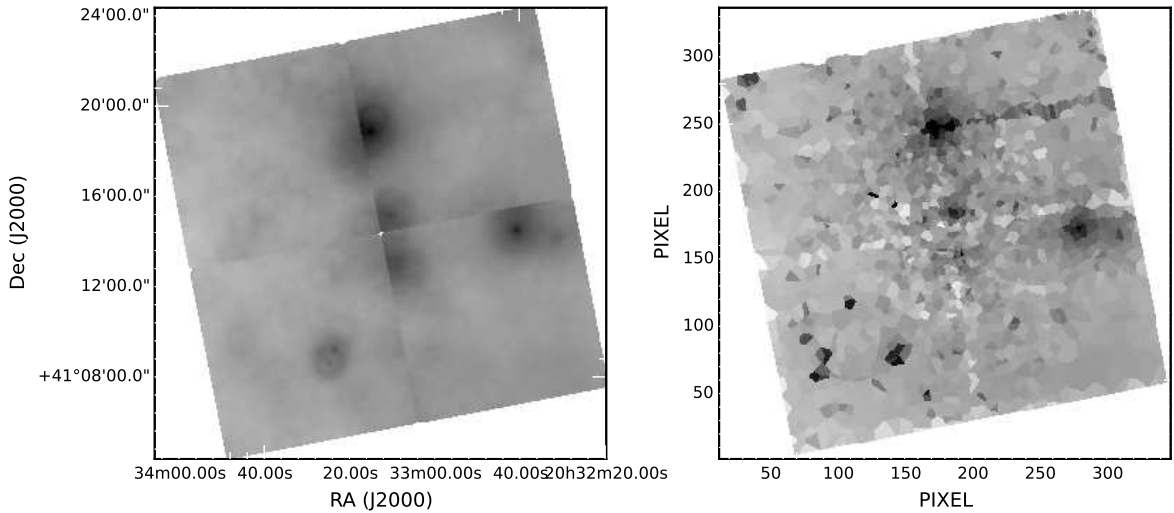}
\caption{Broad-band background maps for ObsID~4511 from the {\it Chandra} Cygnus OB2 Legacy Survey produced by subtracting all detected sources using {\sc wavdetect} (left) and extracted from the properties of the source catalog following our method (right). The color scale is the same in both images and uses a logarithmic scaling that ranges from $7 \times 10^{-4}$ (white) to $7 \times 10^{-2}$ (black) cnts s$^{-1}$ arcmin$^{-2}$. Differences between the two images are due to the differences between {\sc wavdetect} and AE's methods for extracting the local background, correcting for emission from the PSF wings of nearby bright stars \citep[see the {\it better backgrounds} method employed by AE,][]{broo02}, and calculating reliable background estimates.}
\label{background_comparison}
\end{center}
\end{figure*}

For each ObsID we then extract the empirical background count rate at that position, using the background count rates estimated by {\em ACIS Extract} \citep[AE,][]{broo02,broo10} during our source extraction process \citep{wrig14c}. This is possible because our observations contain a high source number density, $\sim$8000 $deg^{-2}$, allowing us to sample the local background by interpolating between nearby extracted sources. Comparisons between background maps made by this method and the more traditional method of subtracting detected point sources from an X-ray image \citep[e.g., using CIAO\footnote{{\it Chandra} Interactive Analysis of Observations, http://cxc.harvard.edu/ciao/} {\sc wavdetect},][]{frus06} show that the two methods produce very similar results (see Figure~\ref{background_comparison}), with the only deviations arising from the different methods AE uses to extract and calculate the background.

\begin{figure*}
\begin{center}
\includegraphics[height=500pt, angle=270]{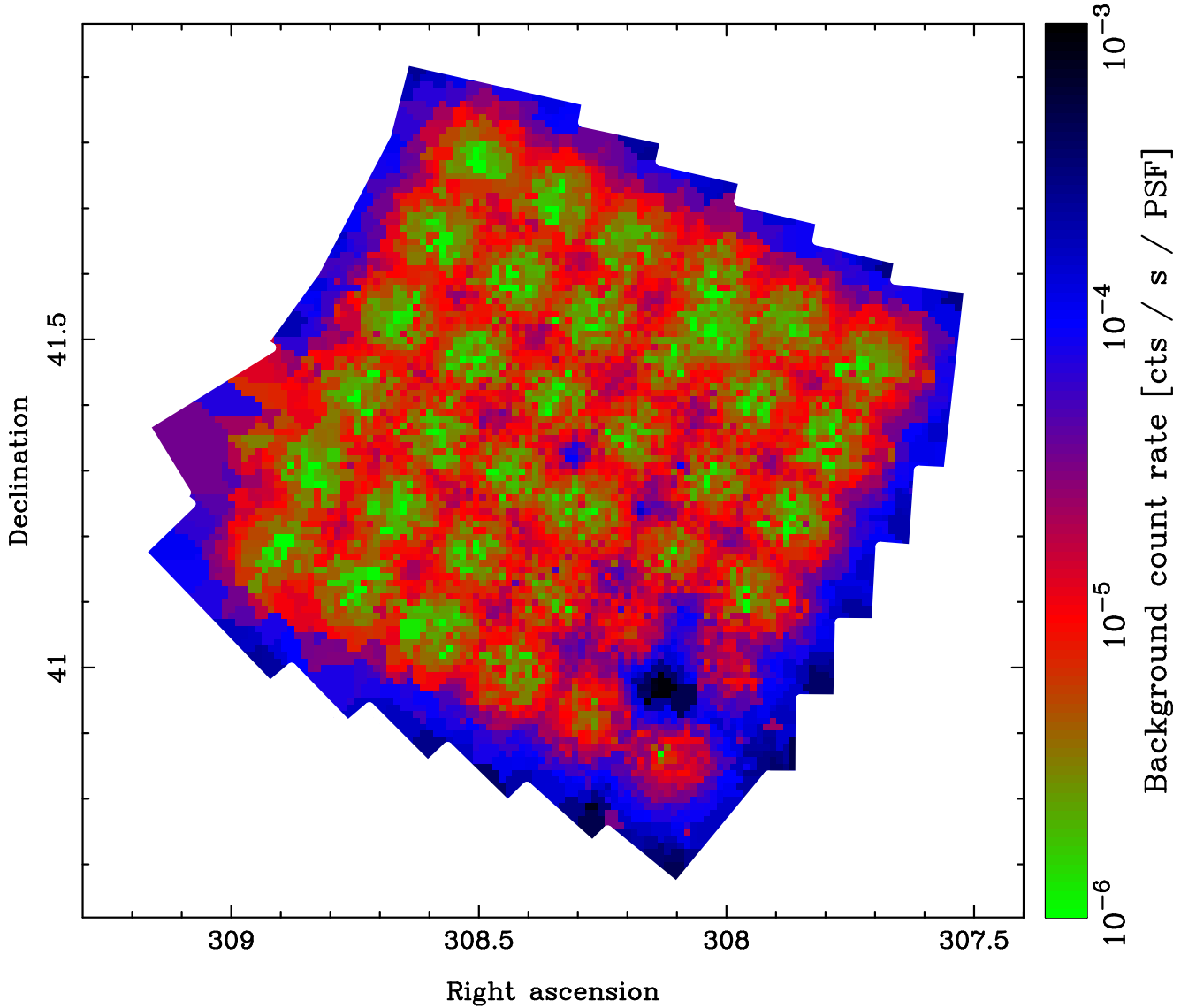}
\caption{Broad-band background map for the {\it Chandra} Cygnus OB2 Legacy Survey extracted from the properties of the source catalog. The background shown is equivalent to $C^b (A^s / A^b)$, or the background count rate per source PSF (see text for more information). For sources observed in multiple ObsIDs we show the lowest background level of all observations for illustrative purposes. The resulting background map shows the tiling strategy adopted by the survey, which leads to a low background level over a large area.}
\label{background_map}
\end{center}
\end{figure*}

We combine the measured background count rate, $C^b$, with the background scaling factor \citep[the ratio of source aperture size to background area, $A^s / A^b$ provided by AE, see Table~2 in][]{wrig14c} to calculate the quantity $C^b (A^s / A^b)$. This is the relevant quantity when calculating the background contribution to the count rate measured in the source aperture and is best described as the background count rate per source PSF (since it also scales with the size of the source extraction aperture, $A^s$, which is equivalent to the source PSF). Figure~\ref{background_map} shows the minimum value of the background count rate per source PSF across our survey area. For regions of the survey observed by more than one ObsID (which is applicable to the majority of our survey area), this quantity is different for each observation (as the local background, $C^b$, and the PSF size, $A^s$, varies with position and off-axis angle). Figure~\ref{background_map} shows the minimum value of this quantity at each point in our survey because this can be a limiting factor in determining whether a source is detected and validated or not. The tiling strategy employed by the survey is evident, as is the high background region surrounding Cyg~X-3 in the south-western corner of the survey area.

The advantage of this method (as opposed to using a background event list with the point sources removed) is that the background is determined with exactly the same method as was used for the data reduction and source validation process that produced our source catalog. This is important because the background count rate used by AE is not just dependent on the true underlying background, but also on the area that is used to sample that background. AE requires that a minimum of 100 events are included in the background region for it to be sufficiently well sampled \citep{broo10}. Therefore the area over which the background is sampled will depend on both how bright the background is at that point (which is dependent on the exposure time of the observation) and also on the level of source crowding in the vicinity of the source itself. High levels of source crowding will cause AE to search for valid background regions over an area further and further from the source itself. By using the empirical background count rate extracted during the true source extraction process we are directly emulating that process.

\subsection{The source detection process}
\label{s-detection}

Once the brightness of a source and the surrounding background level are known we can assess the probability that the source would be detected. Since the source detection process depends only on the number of source and background counts and the detection threshold used, the detection probabilities are independent of the source and background spectra, or the specifics of the detector. Because of this we can parameterize the source detection probability using a set of simulated observations, and then use those source detection probabilities in our Monte Carlo simulation to determine whether a given source is detected.

While our actual source detection procedure employed an array of different methods \citep[CIAO {\sc wavdetect} \citep{free02}, {\it PWdetect} \citep{dami97}, an enhanced multi-ObsID version of {\sc wavdetect} \citep{wrig14c}, as well as employing lists of previously known sources,][]{wrig14c} for simplicity we will only quantify the detection probability from {\sc wavdetect}. Whilst this will slightly under-estimate our sensitivity to sources with a given property, it will provide a reasonable first order estimate of our completeness \citep[the lists of previously known sources did not significantly contribute to the number of detected sources,][]{wrig14c}.

\subsubsection{Source detection simulations}

\begin{figure}
\begin{center}
\includegraphics[height=240pt]{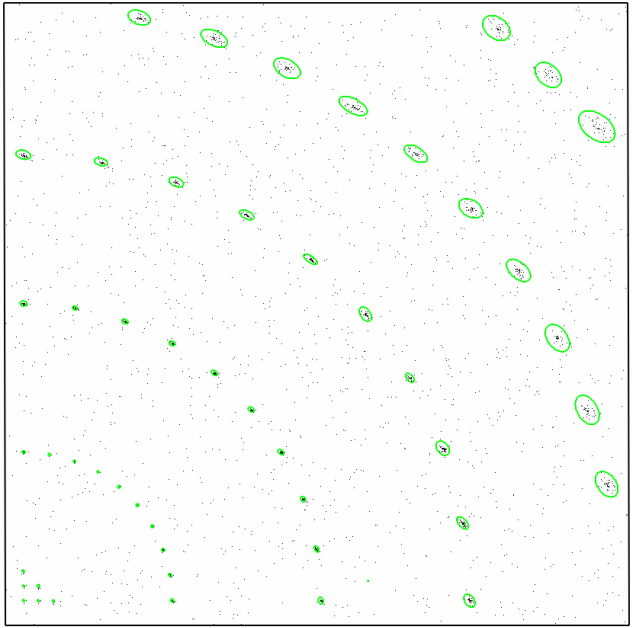}
\caption{Image showing the arrangement of simulated sources on CCD3 of {\it Chandra}'s ACIS detector. The sources are arranged in groups at off-axis angles of 0$^\prime$ (six sources in the bottom left-hand corner at {\it Chandra}'s aimpoint), 2$^\prime$, 4$^\prime$, 6$^\prime$, 8$^\prime$ (ten sources at each off-axis angle arranged in concentric circles) and 10$^\prime$ (three sources in the top right-hand corner). X-ray events (photons) are shown in black and the detected sources (for this example simulation, for which all sources were detected) are marked using green ellipses that illustrate the size enclosing 99.7\% (at 1.49~keV) of {\it Chandra}'s PSF at each off-axis angle.}
\label{source_arrangement}
\end{center}
\end{figure}

To quantify the probability of detecting a given source in our survey we have simulated the detection process to calculate the source detection probability as a function of various parameters. The detection probabilities were calculated by simulating observational data sets using MARX version 5.0.0 \citep{wise03,davi12}. We simulated flat-spectrum sources with 10 intensities from 2 to 1024 counts in steps of $\times$2 counts (this was later complemented by adding extra simulations between the existing steps in the range when the detection probability was $0.05 < P < 0.95$, leading to steps of $\times$$\sqrt{2}$ in these important ranges). We simulated background surface brightness levels at 11 steps from 0.00072 to 0.768 counts pixel$^{-1}$ (in approximate steps of $\times$2 counts pixel$^{-1}$, again with a flat spectrum). MARX simulates sources by distributing their source counts according to the local shape of {\it Chandra}'s PSF at an energy of 1.49~keV, accurately simulating what would be seen in a real observation. The sources were arranged in a fixed pattern across CCD3 of the ACIS detector, at off-axis angles between 0 and 10$^\prime$ at intervals of 2$^\prime$, as illustrated in Figure~\ref{source_arrangement}.

Each simulated data set was processed in the same way as the actual observations; the data were processed using CALDB version 4.5.8 and source detection was performed at scales of 2.0, 4.0, 8.0, 16.0, and 32.0 pixels using {\sc wavdetect} version 4.5. Source detection was performed with false source detection probability thresholds of $10^{-6}$, $10^{-5}$, and $10^{-4}$.

This setup lead to an initial total of 10 source intensities $\times$ 11 background intensities = 110 different simulation configurations. So that our results were not dominated by Poisson uncertainties, we simulated a minimum of 100 sources at each combination of source intensity, background intensity, and off-axis angle. Due to the limited area of the CCD at an off-axis angle of 10$^\prime$ and the large size of the PSF at that distance we were only able to simulate 3 such sources for each CCD simulated. To reach the desired number of 100 simulated sources for each combination of parameters we thus had to simulate at least 34 CCDs for each of the 110 different simulation configurations, resulting in $\sim$3700 individual simulations (additional simulations to better sample the source intensity parameter space brought this total to $\sim$5000 MARX simulations).

We then calculated the detection probability as the fraction of sources detected for each combination of parameters. We did not apply any additional source significance criteria in our simulations \citep[e.g.,][]{zeza07} as this was not applied to the results of our actual source detection process, in which source verification was applied separately (see Section~\ref{s-verification}). Since the detection efficiency at a given threshold is only a function of the number of source and background counts in the detector cell (though the cell size is a function of the off-axis angle), the detection probabilities calculated in this way are independent of the energy band used and can therefore be applied to source detection in any band.

\begin{figure}
\begin{center}
\includegraphics[height=240pt, angle=270]{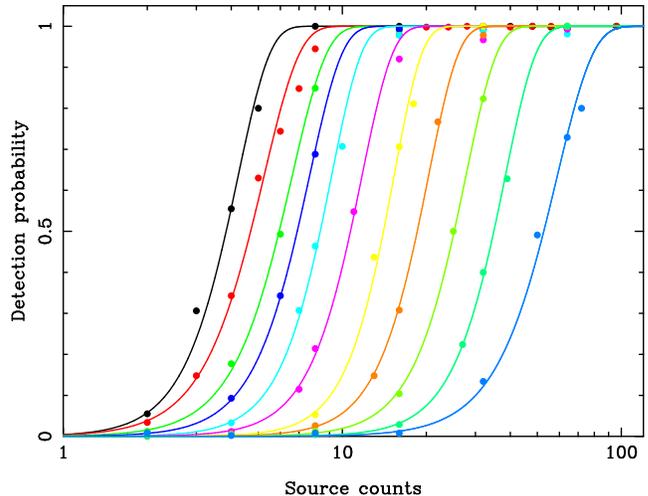}
\caption{Source detection probability as a function of the number of source counts, calculated from the results of a large suite of MARX simulations. All sources were positioned at an off-axis angle of 6$^\prime$ and detected using CIAO {\sc wavdetect} at a threshold of $10^{-5}$. The dots show the results of the source detection simulations, which were performed for various source count intensity levels, while the lines show best fitting detection curves of the form $P = 1 - \mathrm{exp} \left[ \frac{-C^{\lambda_1}}{10^{\lambda_2}} \right]$ (Eqn.~\ref{detection_function}), where the constants $\lambda_1$ and $\lambda_2$ are provided in Table~\ref{parameters}. Each colour represents a different background level, which are (from left to right): 0.00072, 0.0014, 0.0030, 0.0060, 0.012, 0.024, 0.048, 0.096, 0.192, 0.384, and 0.768 counts / pixel.}
\label{detection_curves}
\end{center}
\end{figure}

For each combination of background intensity, off-axis angle, and source detection threshold the detection probability was quantified as a function of the source intensity (in counts), as shown in Figure~\ref{detection_curves}. We fitted each detection probability curve (for each combination of parameters) by a function of the form

\begin{equation}
P = 1 - \mathrm{exp} \left[ \frac{-C^{\lambda_1}}{10^{\lambda_2}} \right]
\label{detection_function}
\end{equation}

\noindent
where $P$ is the detection probability and $C$ is the source intensity (in counts). The constants $\lambda_1$ and $\lambda_2$ (and their uncertainties) were determined for each combination of parameters by maximising the likelihood function for the fit of the model to the data. To determine the highest likelihood values of the parameters $(\lambda_1, \lambda_2)$ we employed the {\tt emcee} Markov Chain Monte Carlo (MCMC) ensemble sampler \citep{fore13}. This method has the advantage of efficiently exploring the parameter space and avoiding local maxima. Our MCMC walks were run for sufficient autocorrelation time so as to ensure a stable distribution of parameters.

\begin{table}
\begin{center}
\caption{Parameters for the source detection probability function} 
\label{parameters}
\begin{tabular}{lllcccc}
\hline
Background	& $\theta$			& Detection	& $\lambda_1$	& $\sigma_{\lambda_1}$	& $\lambda_2$	& $\sigma_{\lambda_2}$ 	\\
(cnts/pixel)	& (arcmin)			& threshold	&\\
\hline
0.00072			& 0				& $10^{-6}$	& 3.61	& 0.21	& 1.70	& 0.07	\\
0.0014			& 0				& $10^{-6}$	& 3.35	& 0.14	& 1.64	& 0.06	\\
0.0030			& 0				& $10^{-6}$	& 2.93	& 0.27	& 1.48	& 0.12	\\
0.0060			& 0				& $10^{-6}$	& 5.74	& 0.55	& 3.80	& 0.34	\\
0.012			& 0				& $10^{-6}$	& 5.99	& 0.61	& 4.27	& 0.40	\\
0.024			& 0				& $10^{-6}$	& 4.93	& 0.52	& 3.89	& 0.36	\\
0.048			& 0				& $10^{-6}$	& 4.87	& 0.36	& 4.19	& 0.30	\\
0.096			& 0				& $10^{-6}$	& 5.24	& 0.56	& 4.89	& 0.50	\\
0.192			& 0				& $10^{-6}$	& 4.76	& 0.60	& 4.99	& 0.58	\\
0.384			& 0				& $10^{-6}$	& 4.68	& 0.48	& 5.32	& 0.52	\\
0.768			& 0				& $10^{-6}$	& 4.86	& 0.79	& 6.08	& 0.95	\\
\hline
\end{tabular} 
\end{center}
{\sc Notes.} The parameters $\lambda_1$ and $\lambda_2$ are used in the parameterised detection probability curve $P = 1 - \mathrm{exp} \left[ \frac{-C^{\lambda_1}}{10^{\lambda_2}} \right]$ (Eqn.~\ref{detection_function}) and were determined by fitting the results of the source detection simulations using an MCMC sampler. 
\newline (This table is available in its entirety in a machine-readable form in the online journal. A portion is shown here for guidance regarding its form and content.)
\newline
\end{table}

The parameterisation of the source detection probability functions as analytic curves has the advantage of smoothing the statistical noise due to the finite number of simulated sources. The parameters for this function are listed in Table~\ref{parameters} (with uncertainties) as a function of the background intensity, off-axis angle, and source detection threshold. The uncertainty on $P$ resulting from using this equation and parameters is estimated to be of the order of $\sim$10\% based on Poisson statistics from the source detection simulations and the uncertainties of the fitting process.

Using the results of these simulations we can quantify the source detection probability for each observation of each source in our Monte Carlo simulation (based on its simulated brightness and the empirical background count rate at that position) and thus determine whether that source would be detected. If the source is detected it is passed onto the source verification process, otherwise it is discarded.

\subsection{The source verification process}
\label{s-verification}

Finally, once a source has been detected its validity is assessed by AE before being included in the final catalog of X-ray sources \citep{wrig14c}. The validity of sources is assessed by testing the null hypothesis that the source does not exist, i.e., that all the events in the source aperture are background events. The probability of this, $P_B$, can be calculated according to the method described by \citet[][Appendix A2]{weis07}. AE calculates $P_B$ under the assumption that $C^b$ is large and therefore that the background is accurately estimated \citep[][Appendix B]{broo10}, under which assumption the expression for $P_B$ approaches the integral of the Poisson distribution over the interval $[C^s, \infty]$

\begin{displaymath}
P_B \simeq 1 - \sum_{i=0}^{C^s-1} \mathrm{Poisson} \left( i \, ; \, (A^s / A^b) C^b \right)
\end{displaymath}

\noindent
where $C^s$ and $C^b$ are the number of counts observed in the source aperture and background regions in a given energy band, and $A^s$ and $A^b$ are the areas of the source aperture and background regions. We imposed a validity threshold of $P_B \leq 0.01$ \citep[as recommended by][]{broo10} for sources in our catalog, which we will also impose in our Monte Carlo simulation of our source verification process.

The number of counts in the source aperture, $C^s$, can be calculated as the sum of the source count rate and the local background count rate, multiplied by the exposure time at the position of the source in each of our observations. The ratio of source to background aperture areas, $A^s / A^b$, and the number of counts in the background region, $C^b$, are dependent on the position of the source in our survey (both are extracted empirically from the observations as described in Section~\ref{s-observations}).

Finally we note that, in calculating $P_B$ for sources observed by more than one different ObsID, AE only considers the combination of observations that minimises $P_B$. This is equivalent to only selecting observations that increase the detection significance of a given source. We adopt the same process in our Monte Carlo simulation by calculating $P_B$ for every combination of observations for a given source and considering only the minimum value. If a source has $P_B \leq 0.01$ in {\it any} such combination of observations we consider the simulated source to have been verified and it would therefore have been included in our catalog.

\subsection{Crowding}

To reproduce the effects of source crowding and confusion on our simulated sources we consider the positions of all simulated sources relative to those of existing `real' sources in our observations. If a simulated source is fainter than its real neighbour and it falls within a distance of twice the radius of the PSF that encloses 40\% of the PSF power at that position then we consider this source to be too close to the existing source and automatically treat it as undetected.  This level was chosen because when AE (which we found to be more conservative than {\sc wavdetect} on these matters) is presented with closely-spaced sources it is prepared to shrink the size of the extraction apertures to a minimum of the 40\% power level before it dismisses one of the sources. When closely-spaced sources are captured by multiple observations we apply this test only to the observation where the sources have the smallest off-axis angle (and therefore smallest PSF), since these observations would be the most influential in separating them. If the simulated source is brighter than the existing source then it is likely that the simulated source would have been detected over the real source and we do not apply this test.

The effects of crowding on the completeness of our observations was found to be very small, influencing the final detection fractions by only $\sim$0.5-1\%. This is due to the fact that even in the centre of the OB association, Cyg~OB2 is not particularly dense \citep{wrig14b}. The area covered by the on-axis PSFs (at 90\% power) of all $\sim$8000 sources detected in our survey represents $<$1\% of the entire survey area, meaning that the probability of a simulated source falling close enough to an existing source to suffer from confusion effects is equally small.

\section{Results}

In this section we present the results of our Monte Carlo simulations based on the model described above. We study the completeness of our observations as a function of each observational (X-ray count rate) and stellar (stellar X-ray luminosity and stellar mass) parameter separately, performing a unique Monte Carlo simulation for each parameter. For each Monte Carlo simulation we performed 1,000,000 draws at each of $\sim$20--40 intervals between the levels of 0\% and 100\% completeness, thus fully sampling the sensitivity curve of our observations.

For each parameter we have calculated the completeness of our observations over both the entire survey area and the central 0.5~deg$^2$ where the total exposure is $\geq$120~ks \citep[see Figure~1 of][]{wrig14c}. For the latter we use a contiguous approximately square region that includes some small areas within this region with exposures $<$120~ks due to chip gaps and misaligned pointings (which are accounted for in our Monte Carlo simulation). This choice reflects the possible use by an observer of sources in a predefined and contiguous area in the center of Cyg~OB2. For each simulation we assume the sources are randomly distributed across the survey area, though we make a small adjustment to this in the final simulation, as explained in Section~\ref{s-cluster}.

\subsection{Completeness results as a function of X-ray source count rate}
\label{s-countrate}

The simplest quantity with which to simulate our completeness is the source count rate. We performed a Monte Carlo simulation for X-ray count rates in the range $10^{-5}$--$10^{-3}$~cnts/s in steps of 0.05~dex, with 1,000,000 randomly-positioned sources simulated at each step. The source count rate provides the intensity of each source in our observations and thus allows us to calculate the source detection probability (using the detection probability curves calculated in Section~\ref{s-detection}) and the source verification probability (as outlined in Section~\ref{s-verification}).

\begin{figure}
\begin{center}
\includegraphics[height=240pt, angle=270]{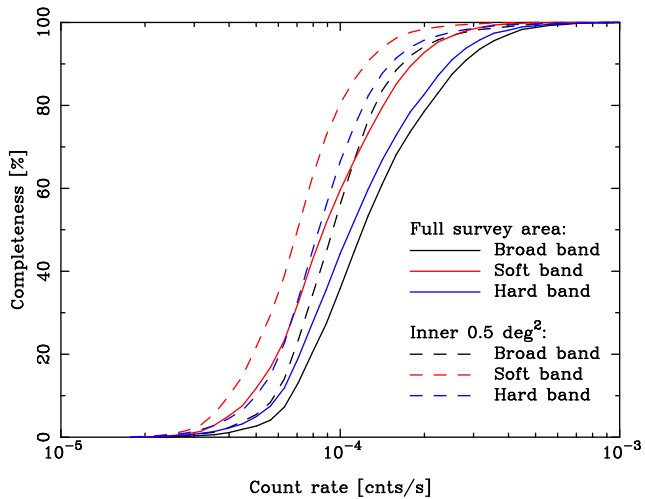}
\caption{Survey completeness as a function of the X-ray source count rate in the {\it broad} (0.5--7~keV, black), {\it soft} (0.5--2~keV, red), and {\it hard} (2--7~keV, blue) bands. The full lines show the completeness over the entire 1~deg$^2$ area, while the dashed lines shows the completeness in the inner 120~ks deep 0.5~deg$^2$ area.}
\label{countrate_sensitivity}
\end{center}
\end{figure}

Figure~\ref{countrate_sensitivity} shows the simulation results as a function of source count rate in each of three energy bands used in our survey: {\it broad} (0.5--7~keV), {\it soft} (0.2--2~keV), and {\it hard} (2--7~keV). For each energy band the background count rate and ratio of extraction aperture areas were extracted from the observations, since both of these quantities can vary with the energy band (e.g., if the background count rate is lower in a particular band then AE must extract events over a larger area to satisfy its requirements on the background).

In the broad band over the entire survey area the completeness increases from 50\% at approximately $1 \times 10^{-4}$ cnts/s to 90\% at $3 \times 10^{-4}$ cnts/s. The exact form of the completeness curve is not smooth and is caused by the unique mixture of different exposure levels in our survey \citep[see e.g. Figure~2 of][]{wrig14c}. In the hard and soft bands the typical completeness is higher at a given count rate (because the background count rate is smaller in a narrower energy band), or equivalently a given completeness level is reached at a lower count rate. The completeness in the inner 0.5~deg$^2$ area is shifted to lower count rates because of the higher exposure in this region, and the sensitivity curve is also notably steeper than that of the entire survey area because of the more uniform exposure level. In the inner 0.5~deg$^2$ area the completeness increases from 50\% at $9 \times 10^{-5}$ cnts/s to 90\% at $1.7 \times 10^{-4}$ cnts/s.

These count rate levels can be attributed to the typical exposure levels of our survey. In the outer 0.5~deg$^2$ the exposure is typically 60~ks, which for a source with a count rate of $6 \times 10^{-5}$~cnts/s (approximately the 10\% completeness level) would result in $\sim$3--4 counts, which is about the minimum number of counts for which a source could be detected on-axis. The 90\% completeness level is reached at about $3 \times 10^{-4}$ cnts/s in the full survey area and $1.7 \times 10^{-4}$ cnts/s in the central 0.5~deg$^2$. These count rates are both equivalent to sources with $\sim$18--20 net counts in our survey (assuming the limiting exposure times in the full survey area and central 0.5~deg$^2$ area are 60 and 120~ks), which therefore represents our completeness level in net counts. This is in good agreement with \citet{broo11} who find a similar completeness limit of $\sim$20 net counts in the {\it Chandra} Carina Complex Project. \citet{broo11} find that their completeness limit is dominated by the detection of sources at large off-axis angles, which we believe is also the main factor in our completeness limit.

\subsection{Completeness as a function of stellar X-ray luminosity}
\label{s-xrayluminosity}

The completeness as a function of stellar X-ray luminosity was calculated by assuming both a distance to the Cyg~OB2 association and a typical spectral shape for PMS stars. Using the X-ray spectrum and luminosity we calculate the count rate in each of the three bands. The sources are then randomly positioned across the survey area and the number of counts is then calculated based on the exposure time of each of the relevant observations. The probability that the source is detected is then calculated in the same way as above. A source only had to pass the source detection test in one of the three energy bands to be considered detected. We performed these simulations between X-ray luminosities of $10^{29}$ and $10^{31}$~erg~s$^{-1}$ in steps of 0.05~dex, with 1,000,000 randomly positioned sources at each step.

\begin{figure}
\begin{center}
\includegraphics[height=240pt, angle=270]{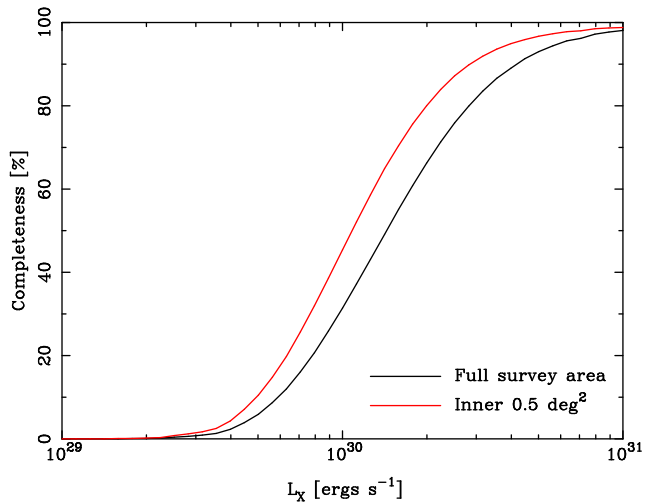}
\caption{Survey completeness as a function of stellar X-ray luminosity for sources across the {\it Chandra} Cygnus~OB2 Legacy Survey area. All sources were modelled at a distance to Cyg~OB2 of 1.4~kpc, with source verification assessed in all three energy bands. The black line shows the completeness over the entire 1~deg$^2$ area, while the red line shows the completeness in the inner 120~ks deep 0.5~deg$^2$ area.}
\label{lx_sensitivity}
\end{center}
\end{figure}

Figure~\ref{lx_sensitivity} shows the completeness of the {\it Chandra} Cygnus~OB2 Legacy Survey as a function of X-ray luminosity, with a rise in completeness from 50\% at $1.4 \times 10^{30}$~ergs~s$^{-1}$ to 90\% at $4 \times 10^{30}$~ergs~s$^{-1}$ over the entire survey area. In the central 0.5~deg$^2$ area the rise in completeness is steeper, going from 50\% completeness at $10^{30}$~ergs~s$^{-1}$ to 90\% at $2.8 \times 10^{30}$~ergs~s$^{-1}$.

\subsection{Completeness as a function of stellar mass}
\label{s-stellarmass}

Completeness as a function of stellar mass was calculated by assuming a relationship between stellar mass and X-ray luminosity. We performed these simulations between stellar masses of 0.1~M$_\odot$ and 3~M$_\odot$ in steps of 0.05~dex, with 1,000,000 randomly positioned sources at each step.

\begin{figure}
\begin{center}
\includegraphics[height=240pt, angle=270]{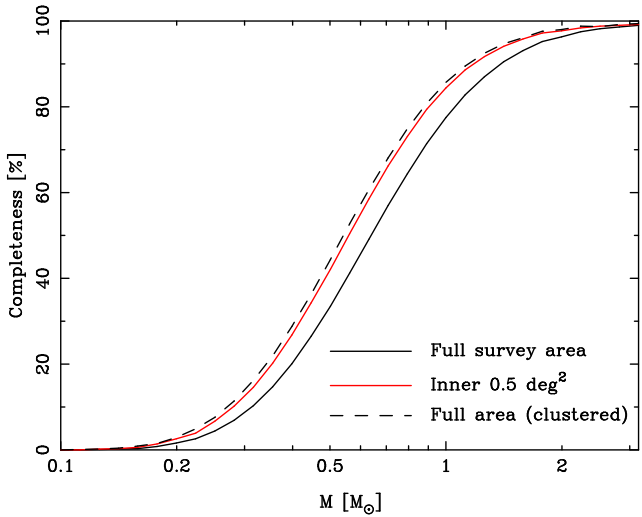}
\caption{Survey completeness as a function of stellar mass for sources across the {\it Chandra} Cygnus~OB2 Legacy Survey area. All sources were modelled at a distance to Cyg~OB2 of 1.4~kpc, with sources verification assessed in all three energy bands. The black line shows the completeness over the entire 1~deg$^2$ area, while the red line shows the completeness in the deep 0.5~deg$^2$ area. The black dashed line shows the completeness of our observations over the entire survey area assuming that the sources are not randomly distributed, but are clustered in the same way as that of the massive stars in Cyg~OB2.}
\label{mass_sensitivity}
\end{center}
\end{figure}

Figure~\ref{mass_sensitivity} shows the completeness of our survey as a function of stellar mass. The results for the entire survey area show a steady rise in completeness with 50\% completeness at 0.6~M$_\odot$ and 90\% at 1.3~M$_\odot$. In the central 0.5~deg$^2$ area the completeness is 50\% at 0.55~M$_\odot$ and 90\% at 1.1~M$_\odot$.

We cannot reliably calculate the the completeness of our observations to intermediate-mass stars ($M > 3 M_\odot$), because it is not clear if intermediate-mass pre-MS stars can generate the stellar magnetic dynamo necessary to produce a high temperature corona that emits X-rays \citep[e.g.,][]{drak14a} or whether the observed X-ray emission from these stars is due to an unresolved binary companion. These questions are explored in more detail using data from the survey by \citet{drak15a}.

\subsection{Completeness of a clustered population as a whole}
\label{s-cluster}

These simulations consider the distribution of sources across our survey area to be random, i.e., we are calculating the completeness of our survey to a  source with a given property that we might detect. However, if we wish to calculate the overall completeness level of our survey then we must account for the fact that the young stars of Cyg~OB2 are not evenly distributed but are centrally concentrated, with a larger fraction of sources in the center of our observations (where the exposure time is longest) and fewer sources at the edges (where the exposure time is shortest).

To account for this we have repeated our simulations as a function of stellar mass but changed the spatial distribution from a random distribution to a centrally concentrated distribution. We model the spatial distribution as a 2-dimensional Gaussian centred at 20:33:00, +41:19:00 with standard deviations of 0.24 and 0.19 degrees in RA and Dec respectively. These parameters were calculated from the spatial distribution of OB stars in the association \citep{wrig15a}, a distribution that is believed to be both relatively large and complete and also representative of the entire population \citep[there is no evidence for mass segregation in Cyg~OB2 indicating that the high and low-mass stars have similar spatial distributions,][]{wrig14b}.

The results of this simulation are illustrated in Figure~\ref{mass_sensitivity} for the entire survey area, showing a completeness of 50\% at 0.55~M$_\odot$ and 90\% at 1.1~M$_\odot$. These results are very similar to those of the inner 0.5~deg$^2$ area for a random distribution of sources (Section~\ref{s-stellarmass}). This is because the clustered spatial distribution of sources produces 87\% of sources within the central 0.5~deg$^2$ area of the survey, compared to $\sim$50\% for the random distribution of sources. Since the majority of sources in Cyg~OB2 are within the central 0.5~deg$^2$ area, that area contributes the most to the overall completeness of the survey to a clustered population.

\section{Summary}

We have used a hierarchical Monte Carlo simulation to calculate the completeness of a complex arrangement of X-ray observations to an underlying population of X-ray emitting sources with given properties. The sensitivity of the observations is determined by calculating the fraction of sources with a given property that are both detected and verified in the same way as those used to process the data. These simulations are built on the empirical properties of the observations including the background level across all the individual observations, an analytical estimate of the source verification procedure, and an extensive set of simulations of the source detection process. Both the source detection simulations and the overall detection and verification Monte Carlo simulation can be easily adapted for use with other X-ray observations and surveys. These simulations could be extended in the future to assess the products of the source detection and verification process with respect to the input parameters, allowing issues such as Eddington bias to be quantified.

We have used these simulations to calculate the completeness of the {\it Chandra} Cygnus~OB2 Legacy Survey observations to an underlying stellar population, as a function of various observational and stellar parameters. We find that the survey reaches a 90\% completeness level for a PMS population at the distance of Cyg~OB2 at an X-ray luminosity of $3 \times 10^{30}$~ergs~s$^{-1}$ and a stellar mass of 1.3~M$_\odot$. When considering only the inner 0.5~deg$^2$ of the survey with the deepest observations we find that the survey reaches a 90\% completeness level at an X-ray luminosity of $2.8 \times 10^{30}$~ergs~s$^{-1}$ and a stellar mass of 1.1~M$_\odot$. We also show that when considering the underlying population to be clustered and not randomly distributed we find that our survey reaches a 90\% completeness level over the entire area of the observations at 1.1~M$_\odot$.

\acknowledgments

The authors would like to thank the anonymous referee for a prompt and helpful referee's report that improved the content and clarity of this paper. The authors would also like to thank Salvatore Sciortino for advice and comments on this paper. NJW acknowledges a Royal Astronomical Society Research Fellowship. NJW and MG acknowledge support from {\it Chandra} grant GO0-11040X, and MG acknowledges support from grant PRIN-INAF~2012 (P.I.~E.~Flaccomio). JJD and VLK were supported by NASA contract NAS8-03060 to the {\em Chandra X-ray Center} during the course of this research and thank the Director, B.~Wilkes, for continuing support. This research has made use of data from the {\it Chandra X-ray Observatory} and software provided by the {\it Chandra X-ray Centre} in the application packages CIAO and Sherpa, and from Penn State for the {\sc ACIS Extract} software package.

\bibliographystyle{apj}

\end{document}